\documentstyle[11pt]{article}

\newcommand{\disregard}[1]{}

\disregard{
\textwidth       6.5  in
\textheight      9.5  in
\oddsidemargin   0    in
\evensidemargin  0    in
\topmargin      -0.5  in

\baselineskip 12.5 pt

\footnotesep=1.4\footnotesep
}

\def\topfraction{1.0}
\def\bottomfraction{1.0}

\newcommand{\be}{\begin{equation}}
\newcommand{\ee}{\end{equation}}
\newcommand{\ba}{\begin{array}}
\newcommand{\ea}{\end{array}}
\newcommand{\bt}{\begin{tabular}}
\newcommand{\et}{\end{tabular}}
\newcommand{\bc}{\begin{center}}
\newcommand{\ec}{\end{center}}

\newcounter{leteq}
\newcommand{\steplet}{\stepcounter{leteq}\addtocounter{equation}{-1}}

\newenvironment{eqnalpha}{\setcounter{leteq}{1}

\begin{eqnarray}}{\end{eqnarray}%
}

\newenvironment{eqnalphalabel}[1]{\setcounter{leteq}{1}
\raisebox{0cm}[0cm][0cm]{\begin{minipage}{1cm}%
\begin{eqnarray}\label{#1}&&\nonumber\end{eqnarray}\end{minipage}}

\begin{eqnarray}}{\end{eqnarray}%
}

\newcommand{\bn}{\begin{eqnalpha}}
\newcommand{\en}{\end{eqnalpha}}
\newcommand{\bmln}[1]{\begin{eqnalphalabel}{#1}}
\newcommand{\emln}{\end{eqnalphalabel}}

\newcommand{\bbox}[1]{\mbox{{\boldmath{$#1$}}}}
\newcommand{\text}[1]{\mbox{\footnotesize{#1}}}
\newcommand{\tensor}[1]{\stackrel{\leftrightarrow}{#1}}
\newcommand{\bboxr}{\bbox{r}}

\newcommand{\spc}{{\ }}

\newcommand{\jde}{{\cal J}^{(2)}}
\newcommand{\umi}{$\hbar^2\!$/MeV}
\newcommand{\ptzzu}{$\pi[301]1/2(r$=$+i$)}

\begin{document}

\vspace*{23mm}
\centerline{\Large\bf TIME-ODD COMPONENTS IN THE ROTATING}
\medskip
\centerline{\Large\bf MEAN FIELD AND IDENTICAL BANDS$^{*,**}$}

\def\thefootnote{*}
\footnotetext[1]{Presented at
the XXIII Mazurian Lakes School on Physics,
Piaski, Poland, August 23 -- September 3, 1995.}%
\def\thefootnote{**}
\footnotetext[2]{This research was supported in part
by the Polish Committee for Scientific Research
under Contract No.~2P03B~034~08.
Numerical calculations were performed
at the
{\it Institut du D\'eveloppement et de Ressources en Informatique
Scientifique} (IDRIS) of CNRS, France, under Project No.~940333.
}

\vskip 5 mm

\centerline{\large\sc  \underline{J. Dobaczewski}}

\vskip 3 mm

\centerline{   Institute of Theoretical Physics, Warsaw University}
\centerline{         Ho\.za 69, PL-00-681 Warsaw, Poland}

\vskip 3 mm

\centerline{\large\sc  and}

\vskip 3 mm

\centerline{\large\sc  J. Dudek}

\vskip 3 mm

\centerline{   Centre de Recherches Nucl\'eaires,
               IN$_2$P$_3$--CNRS/Universit\'e Louis Pasteur}
\centerline{   F-67037 Strasbourg Cedex 2, France}

\vskip 8 true mm

\begin{center}
\parbox{11.1cm}{
\noindent
\small
A systematic construction of the energy-density functional
within the local density approximation is presented.  The
Hartree-Fock equations corresponding to such a functional are
solved in case of rotating superdeformed nuclei. The identical
bands in $^{152}$Dy, $^{151}$Tb, and $^{150}$Gd are
investigated and the time-odd components in the rotating mean
field are analyzed.

\smallskip
\noindent
PACS numbers: 21.60.Ev, 21.10.Re, 21.60.Jz, 27.70.+q
}
\end{center}

\vspace{2mm}

\section{Introduction}
\label{sec1}

The nuclear rotation is an example of a collective motion for
which time-dependent
linear combinations of stationary states of a given spin
can be identified with a rotating wave packet. A description of
such states can be performed witin a mean-field theory in which
properties of the system are determined by the one-body density
matrix.  Because the time-reversal symmetry is broken for the
rotating states, the density matrix, as well as the resulting
mean field have both time-even and time-odd components.

Properties of nuclear time-even mean fields are known rather
well, because they are reflected in multiple static phenomena
which can be studied experimentally, cf.~review \cite{AFN90a}.
On the other hand, very little is known about properties of the
time-odd mean fields. The present study is devoted to an attempt
to analyse properties of these time-odd components in rapidly
rotating superdeformed nuclei. In particular, we aim at studying
the phenomenon of identical bands, which may provide a unique
information on poorly known sector of the nuclear effective
interaction.

In Sect.\spc\ref{sec2} we present a construction of the
energy-density functional based on a few simple assumptions
concerning its structure. Section \ref{sec3} presents results of
selfconsistent cranking method applied to nuclei in the
$A$$\simeq$150 region; it supplements the results of similar
calculations presented in Ref.\spc\cite{J2D2}.

\section{Local density approximation}
\label{sec2}

The density of nuclear matter in the interior of atomic nuclei
has a well defined value, called the saturation density, which
is independent of the nuclear size. This property of matter
composed of strongly interacting nucleons is a basic feature of
the nucleon-nucleon interaction which
(depending also on the intrinsic symmetries of the nucleonic
states) gives the strongest
attraction at a certain distance and becomes weaker or repulsive
at other distances.  Therefore, as a first approximation one
can consider a state of the matter in a given point of the
nucleus as given by that of the infinite matter at the saturation
density.  In other words, properties of nuclear matter in a
given point weakly depend on other points of the nucleus.

Such an observation underlies the local density
approximation (LDA) in nuclear physics \cite{RS80}. It has been
adopted from similar ideas developed in atomic physics
\cite{DG90} and reflects fundamental properties of
many-fermion systems. In such systems, the Pauli correlations
play a very important role in defining properties of states near
its ground state.  To some extent, details of interactions
are less important, and in particular, the local density
approximation can be valid irrespectively of the interaction
range.

Because of the finite nuclear size, the simplest version of the
LDA has to be corrected by taking into account not only the
density itself but also its local derivatives.  This amounts to
extending the Thomas-Fermi approximation beyond its extreme
version, or to applying the systematic Wigner-Kirkwood $\hbar$
expansion \cite{RS80}.  In a phenomenological approach one may
use the ideas of the LDA to construct the energy-density
functionals, and in what follows we proceed along these lines.

Principal assumptions of such a LDA can be formulated as
follows:
\begin{itemize}
\item
The total energy of the nucleus is given
by the integral of the local energy density ${\cal H}(\bbox{r})$,
   \begin{equation}\label{eq107}
   {\cal E} = \int d^3\bbox{r} {\cal H}(\bbox{r}),
   \end{equation}
\item
The energy density depends on the nuclear one-body density matrix and
its derivatives up to the second order.
\end{itemize}

Neglecting for simplicity
the isospin degree of freedom, i.e., considering one type
of nucleons only, the density matrix in spatial coordinates
can be defined by
   \begin{equation}\label{eq301}
   \rho(\bboxr\sigma,\bboxr'\sigma') =
   \langle\Phi| a^+(\bboxr'\sigma') a(\bboxr\sigma)|\Phi \rangle,
   \end{equation}
where $\bboxr$ is the position and $\sigma$ is the spin of a nucleon,
while $|\Phi\rangle$ is a many-body nuclear wave function.

The spin degrees of freedom can be separated by defining the scalar
and the vector parts of the density matrix,
$\rho(\bboxr,\bboxr')$ and $\bbox{s}(\bboxr,\bboxr')$, respectively,
i.e.,
   \bmln{eq302}
   \rho(\bboxr,\bboxr') &=&
    {\displaystyle\sum_\sigma}\rho(\bboxr\sigma,\bboxr'\sigma)
                                                        \\[1ex] \steplet
   \bbox{s}(\bboxr,\bboxr') &=&
    {\displaystyle\sum_{\sigma\sigma'}}\rho(\bboxr\sigma,\bboxr'\sigma')
                        \langle\sigma'|\bbox{\sigma}|\sigma\rangle .
   \emln
Since the density matrix is hermitian, its scalar and vector
parts have the following properties with respect to exchanging
spatial arguments:
   \bmln{eq303}
    \rho^*(\bboxr,\bboxr') &=&
    \rho(\bboxr',\bboxr)
                                                        \\[1ex] \steplet
     \bbox{s}^*(\bboxr,\bboxr') &=&
     \bbox{s}(\bboxr',\bboxr) .
   \emln
On the other hand, the density matrices corresponding to
time-reversed states read:
   \bmln{eq304}
   \rho^T(\bboxr,\bboxr') &=&
   \rho^*(\bboxr,\bboxr')
                                                        \\[1ex] \steplet
   \bbox{s}^T(\bboxr,\bboxr') &=&
   -\bbox{s}^*(\bboxr,\bboxr') .
   \emln
These properties mean that for the time-even state
the scalar density matrix is a real and symmetric function of
spatial arguments, whereas the vector density matrix is
imaginary and antisymmetric. Therefore,
the local parts of the scalar and of the vector densities,
   \bmln{eq305}
   \rho    (\bboxr) &=& \rho    (\bboxr,\bboxr) \\ \steplet
   \bbox{s}(\bboxr) &=& \bbox{s}(\bboxr,\bboxr) ,
   \emln
respectively, do not and do change the sign under the time reversal.
Hence the matter density $\rho(\bboxr)$ is time-even,
and the spin density $\bbox{s}(\bboxr)$ is time-odd.

\begin{table}[t]
\caption[T]{%
Local densities of a fermion system up to the second order
in derivatives with respect to the relative coordinate}
\begin{center}
\begin{tabular}{|l|c|l|}
\hline
Type of density & Order & Defintion \\
\hline
 Matter:          &0& $\rho(\bboxr)=\rho(\bboxr,\bboxr)$ \\
 Current:         &1& $\bbox{j}(\bboxr)= (1/2i)
          [(\bbox{\nabla}$$-$$\bbox{\nabla}')\rho(\bboxr,\bboxr')]_{r=r'}$
                                                         \\
 Kinetic:         &2& $\tau(\bboxr)=
            [\bbox{\nabla}\cdot\bbox{\nabla}'\rho(\bboxr,\bboxr')]_{r=r'}$
                                                         \\
 Spin:            &0& $\bbox{s}(\bboxr)=\bbox{s}(\bboxr,\bboxr)$
                                                         \\
 Spin current:    &1& $J_{\mu\nu}(\bboxr)=(1/2i)
         [({\nabla}_\mu$$-$${\nabla}'_\mu){s}_\nu(\bboxr,\bboxr')]_{r=r'}$
                                                         \\
 Spin kinetic:    &2& $\bbox{T}(\bboxr)=
        [\bbox{\nabla}\cdot\bbox{\nabla}'\bbox{s}(\bboxr,\bboxr')]_{r=r'}$
                                                         \\
\hline
\end{tabular}
\end{center}
\label{tab1}
\end{table}

The local derivatives of the density
matrix have been defined in Ref.\spc\cite{EBG75}
and are summarized in Table \ref{tab1}.
(See Ref.\spc\cite{Pas76} for another possible set of definitions.)
A systematic construction \cite{EBG75}
of these derivatives up to a given order consists
in acting on the scalar and vector density matrices (\ref{eq302}) with
the operators $(\bbox{\nabla}$$-$$\bbox{\nabla}')/2i$
and setting the arguments equal, $\bboxr$=$\bboxr'$, after
the differentiation is completed. This amounts to calculating
derivatives with respect to the relative coordinate
$\bboxr$$-$$\bboxr'$ at the relative coordinate equal to zero.
At this point, the derivatives $(\bbox{\nabla}$+$\bbox{\nabla}')/2$
with respect to the total
coordinate $\bboxr$+$\bboxr'$ are not considered, because they
amount to calculating the derivatives {\em after} setting
$\bboxr$=$\bboxr'$, see below.

In the first order one obtains two new local densities,
the current density $\bbox{j}$ and the spin current density
$J_{\mu\nu}(\bboxr)$, Table \ref{tab1}.
In the second order one has to
act on the scalar and vector densities with the tensor operator
$({\nabla}_\mu$$-$${\nabla}'_\mu)$$({\nabla}_\nu$$-$${\nabla}'_\nu)$.
However, the resulting traceless-symmetric-tensor density
is not interesting, because there is no other zero-order tensor density
with which it could have been contracted
to construct the corresponding term in the
energy density, see below. Therefore,
only the scalar part of the second-order operator should be
considered. Due to the identity
   \begin{equation}\label{eq306}
   (\bbox{\nabla}-\bbox{\nabla}')^2=
   (\bbox{\nabla}+\bbox{\nabla}')^2 - 4\bbox{\nabla}\cdot\bbox{\nabla}',
   \end{equation}
the action of $(\bbox{\nabla}-\bbox{\nabla}')^2$ can be expressed by that
of $\bbox{\nabla}\cdot\bbox{\nabla}'$, and therefore, following
Ref.\spc\cite{EBG75} one uses
kinetic densities shown in Table \ref{tab1}.

We may now proceed by constructing the local energy density as
a sum of terms depending on the above local densities. Such a construction
is based on the following rules:

\begin{itemize}

\item
{\bf Every term should be quadratic in local densities.}

An {\it a priori} arbitrary coupling constant can be used in front of
every term.  It is well known that the energy densities which
are quadratic in the matter density do not lead to saturating
systems, and therefore, a density dependence of the coupling
constants should be allowed.

\item
{\bf The energy density should be invariant with respect to parity,
time reversal, and rotations.}

This invariance should be considered in a limit of unbroken
symmetries. In principle, an extension of the functional beyond
this limit is arbitrary. For example, for broken parity symmetry
one could use in the functional either the term
$C^{\rho}\rho^2$, or two independent terms $C^{\rho^+}\rho_+^2$
and $C^{\rho^-}\rho_-^2$, where $\rho_\pm$=$(\rho\pm\rho^P)/2$,
and where $\rho^P$ corresponds to the parity inversed state. It
is obvious that for the unbroken symmetry both choices give the
same result, provided that $C^{\rho}$=$C^{\rho^+}$.  Moreover,
for both choices the functional is invariant with respect to
parity. On the other hand, for the broken symmetry the first
choice gives the following contribution to the total energy:
   \begin{equation}\label{eq308}
   \int d^3\bboxr C^{\rho}\rho^2 =
   \int d^3\bboxr C^{\rho}(\rho_+^2+\rho_-^2),
   \end{equation}
and therefore it requires that $C^{\rho^-}$=$C^{\rho^+}$, while
in the second choice these two coupling constants can be chosen
independently.  These ambiguities are even larger for
density-dependent coupling constants. Up two now, mostly the
first choice has been considered, i.e., the functionals are
constructed in the unbroken symmetry limit.

\item
{\bf Terms beyond the second order are not taken into account.}

The order of the given term is defined as a sum of orders of
both densities on which this term depends. The requirement of
neglecting the terms beyond the second order is consistent with
not considering the densities beyond the second order, see Table
\ref{tab1}. This amounts to an expansion of the energy density
up to the second order in relative coordinates.

\end{itemize}

\begin{table}[t]
\caption[T]{%
Construction of all possible terms up to second order in the energy
density of a fermion system (for comments see text).}
\begin{center}
\begin{tabular}{|l|l|c|ccc||l|l|}
\hline
 \multicolumn{1}{|c|}{1} & \multicolumn{1}{c|}{2} & 3 &
 \multicolumn{3}{c||}{4} & \multicolumn{2}{c|}{5} \\
\hline
Density & Derivative & Order
                     & \multicolumn{3}{c||}{Symmetry}
                     & \multicolumn{2}{c|}{Energy density} \\
\cline{4-8}
             &               & &   T   &   P  &   space & prime & dependent \\
\hline
 $\rho(\bboxr)            $ &&0&   +   &   +  &  scalar & $\rho^2$  & \\
&$\bbox{\nabla}\rho        $ &1&   +   &  $-$ &  vector & &
                                                  $(\bbox{\nabla}\rho)^2$ \\
&$\Delta\rho               $ &2&   +   &   +  &  scalar & $\rho\Delta\rho$ & \\
&$(\nabla$$\times$$\nabla)^{(2)}\!\rho$
                             &2&   +   &   +  &  tensor &  & \\
 $\tau(\bboxr)            $ &&2&   +   &   +  &  scalar & $\rho\tau$  & \\
 $J^{(0)}(\bboxr)         $ &&1&   +   &  $-$ &  scalar &
                                            $\big(J^{(0)}\big)^2$ & \\
&$\bbox{\nabla}
           J^{(0)}        $  &2&   +   &   +  &  vector & & \\
 $\bbox{J}(\bboxr)        $ &&1&   +   &  $-$ &  vector & $\bbox{J}^2$  &
                                        $\bbox{\nabla}\rho\cdot\bbox{J}$  \\
&$\bbox{\nabla}\cdot
          \bbox{J}        $  &2&   +   &   +  &  scalar & $\rho\bbox{\nabla}
                                                       \cdot\bbox{J}$  & \\
&$\bbox{\nabla}\times
          \bbox{J}        $  &2&   +   &   +  &  vector &  & \\
&$(\nabla$$\times$$J)^{(2)}$ &2&   +   &   +  &  tensor &  & \\
 $J^{(2)}_{\mu\nu}(\bboxr)$ &&1&   +   &  $-$ &  tensor &
                                                   $\big(J^{(2)}\big)^2$ & \\
 $\bbox{s}(\bboxr)        $ &&0&  $-$  &   +  &   vector & $\bbox{s}^2$  & \\
&$\bbox{\nabla}\cdot
          \bbox{s}        $  &1&  $-$  &  $-$ &   scalar &
                                      $(\bbox{\nabla}\cdot\bbox{s})^2$ & \\
&$\bbox{\nabla}\times
          \bbox{s}        $  &1&  $-$  &  $-$ &   vector &   &
                                      $(\bbox{\nabla}\times\bbox{s})^2$  \\
&$(\nabla$$\times$$s)^{(2)}$ &1&  $-$  &  $-$ &  tensor &  &
                               $\big((\nabla$$\times$$s)^{(2)}\big)^2$   \\
&$\Delta\bbox{s}          $  &2&  $-$  &   +  &   vector & $\bbox{s}
                                                 \cdot\Delta\bbox{s}$  & \\
 $\bbox{j}(\bboxr)        $ &&1&  $-$  &  $-$ &   vector & $\bbox{j}^2$  &
                         $\bbox{j}\cdot\bbox{\nabla}\times\bbox{s}$      \\
&$\bbox{\nabla}\cdot
          \bbox{j}        $  &2&  $-$  &   +  &   scalar &  & \\
&$\bbox{\nabla}\times
          \bbox{j}        $  &2&  $-$  &   +  &   vector & $\bbox{s}\cdot
                                         \bbox{\nabla}\times\bbox{j}$  & \\
&$(\nabla$$\times$$j)^{(2)}$ &2&  $-$  &   +  &  tensor &  &               \\
 $\bbox{T}(\bboxr)        $ &&2&  $-$  &   +  &   vector & $\bbox{s}\cdot
                                                            \bbox{T}$  & \\
\hline
\end{tabular}
\end{center}
\label{tab2}
\end{table}

The construction of the energy density is illustrated in Table
\ref{tab2}. The first column gives the densities presented in
Table \ref{tab1}. The spin current density $J_{\mu\nu}$ is split
into the standard scalar, vector, and tensor parts $(J^{(0)}$,
$\bbox{J}$, and $J^{(2)}$, respectively) according to
   \begin{equation}\label{eq309}
   J_{\mu\nu} = \frac{1}{3}J^{(0)} \delta_{\mu\nu}
              + \frac{1}{2}\epsilon_{\mu\nu\kappa}J_{\kappa}
              + J^{(2)}_{\mu\nu} ,
   \end{equation}
where
   \bmln{eq310}
   J^{(0)}    &=& \sum_{\mu}J_{\mu\mu}
                                                    ,  \\ \steplet
   J_{\kappa} &=& \sum_{\mu\nu}\epsilon_{\kappa\mu\nu}J_{\mu\nu}
                                                    ,  \\ \steplet
   J^{(2)}_{\mu\nu} &=&   \frac{1}{2}(J_{\mu\nu}+J_{\nu\mu})
                        - \frac{1}{3}\delta_{\mu\nu}
                          \sum_{\kappa}J_{\kappa\kappa} .
   \emln

The second column of Table \ref{tab2} presents derivatives of
the densities from the first column. The derivatives are
constructed in the vector-coupled form.  For example, from the
(vector) gradient operator $\bbox{\nabla}$ and the scalar
density $\rho$ one can obtain the vector density
$\bbox{\nabla}\rho$, the scalar density $\Delta\rho$, and the
tensor density $(\nabla$$\times$$\nabla)^{(2)}\rho$.  Since
$\rho$ itself is a zero-order density, the density
$\bbox{\nabla}\rho$ is of the first order and the other two
densities are of the second order, as indicated in the
third column of the Table.

The derivatives are calculated only up to the second order.
Moreover, the rank-3 tensors are not taken into account and the
derivatives of $J^{(2)}_{\mu\nu}(\bboxr)$ are not included in
the Table, because there are no other densities with the same
symmetries to construct terms in the energy density.
Apart from these two omissions, the Table presents a complete
set of densities and their derivatives which can be obtained
within the specified limits.

The fourth column of the Table gives the symmetries of
every density in the unbroken symmetry limit, i.e.,
assuming that $\rho$ is a time-even, parity-even scalar density,
and $\bbox{s}$ is a time-odd, parity-even vector density.
Then the symmetries presented in the Table result from the fact
that the gradient operator $\bbox{\nabla}$ is a time-even, parity-odd
vector operator.

Finally, the fifth column of the Table presents the terms in the
energy density which can be constructed from the densities
listed in the first two columns. In each case a term is obtained by
forming a scalar, time-even, and parity-even product of two
densities, i.e., by multiplying densities having exactly the
same symmetries. Two avoid the double-counting, a given density
is multiplied only by densities appearing higher in the Table.
The rule of not including terms obove the second order
is also respected.

For example, according to this construction scheme, the
matter density $\rho$ gives the simplest term in the
energy density, $\rho^2$, as shown in the first line
of the Table. Similarly, the derivative $\bbox{\nabla}\rho$
gives rise to the term $(\bbox{\nabla}\rho)^2$. This term is
listed in the column of ``dependent'' terms, because
for a density-independent coupling constant it can be
expressed by the ``prime'' term $\rho\Delta\rho$ given
in the third line. This can be done by integrating
by parts the integral of the energy density (\ref{eq107}).
The corresponding expressions for five ``dependent''
terms are as follows:
   \bmln{eq307}
   (\bbox{\nabla}\rho)^2           &=& - \rho\Delta\rho
                                                      ,  \\ \steplet
   \bbox{\nabla}\rho\cdot\bbox{J}  &=& - \rho\bbox{\nabla}\cdot\bbox{J}
                                                      ,  \\ \steplet
   (\bbox{\nabla}\times\bbox{s})^2 &=& - (\bbox{\nabla}\cdot\bbox{s})^2
                                       -  \bbox{s}\cdot\Delta\bbox{s}
                                                      ,  \\ \steplet
    \sum_{\mu\nu} \big((\nabla\times s)^{(2)}_{\mu\nu}\big)^2
                      &=&  \frac{2}{3} (\bbox{\nabla}\cdot\bbox{s})^2
                                                      ,  \\ \steplet
    \bbox{j}\cdot\bbox{\nabla}\times\bbox{s} &=&
                          \bbox{s}\cdot\bbox{\nabla}\times\bbox{j} .
   \emln

Altogether, this construction procedure gives eighteen terms in
the energy density, of which thirteen are ``prime'' and five are
``dependent''. Three terms correspond to the squares of the spin
current density $J_{\mu\nu}(\bboxr)$ for the three different
values of the intermediate-coupling angular momentum. Usualy,
these three independent terms are considered together in the
following single term:
   \begin{equation}\label{eq311}
   \tensor{J}^2 = \sum_{\mu\nu} J^2_{\mu\nu}
                = \frac{1}{3}\big(J^{(0)}\big)^2
                + \frac{1}{2}\bbox{J}^2
                + \sum_{\mu\nu} \big(J^{(2)}_{\mu\nu}\big)^2
   \end{equation}
with a common coupling constant.

It turns out that each density, or a derivative of the density,
gives rise to at most one ``prime'' term in the energy density,
and these terms and coupling constants can be conveniently
labelled by the corresponding densities. We then obtain the
following expression for the interaction energy density:
   \begin{equation}\label{eq312}
   {\cal H}^{\text{int}}(\bbox{r})  =  {\cal H}^{\text{even}}(\bbox{r})
                         +  {\cal H}^{\text{odd}} (\bbox{r}) ,
   \end{equation}
with
   \bmln{eq109}
   {\cal H}^{\text{even}}(\bbox{r})
    &=& C^{\rho}            \rho^2
     +  C^{\Delta\rho}      \rho\Delta\rho
     +  C^{\tau}            \rho\tau
     +  C^{   J}            \tensor{J}^2
                                                 \nonumber \\ &&
     +  C^{\nabla J}        \rho\bbox{\nabla}\cdot\bbox{J},
                                                 \label{eq109a}
                                                 \\ \steplet
   {\cal H}^{\text{odd}}(\bbox{r})
    &=& C^{   s}            \bbox{s}^2
     +  C^{\Delta s}        \bbox{s}\cdot\Delta
                              \bbox{s}
     +  C^{   T}            \bbox{s}\cdot\bbox{T}
     +  C^{   j}            \bbox{j}^2
                                                 \nonumber \\ &&
     +  C^{\nabla j}        \bbox{s}\cdot(\bbox{\nabla}\times\bbox{j})
     +  C^{\nabla s}        (\bbox{\nabla}\cdot\bbox{s})^2 .
                                                 \label{eq109b}
   \emln
Each of these terms can have the isoscalar and the isovector
form, and the kinetic energy density has to be added to obtain
the total energy density of a nuclear system.

Apart from the last term in ${\cal H}^{\text{odd}}(\bbox{r})$,
all these terms appear in the energy density corresponding to
the Skyrme interaction \cite{EBG75,J2D2}. The additional term in
Eq.\spc(\ref{eq109b}) can only be obtained if the tensor
interaction is added to the Skyrme force, i.e., when the results
of Ref.\spc\cite{SBF77} are generalized to the case of broken
time-reversal symmetry.  In the following we do not consider such
a possibility, and we set $C^{\nabla s}$=0.

By varying the total energy (\ref{eq107}) with respect to the
single-particle wave functions one obtains the Hartree-Fock
time-even and time-odd mean fields \cite{EBG75}. In the
convention corresponding to the energy density of
Eq.\spc(\ref{eq109}), the relevant expressions are given in
Ref.\spc\cite{J2D2} and will not be repeated here.  Instead, in
the following section we present the results of the Hartree-Fock
cranking method applied to the description of superdeformed
bands.

\section{Superdeformed bands in $^{152}$Dy, $^{151}$Tb, and $^{150}$Gd}
\label{sec3}

In Fig.\spc\ref{fig01} we show the results of calculations for
the yrast superdeformed band in $^{152}$Dy and for the first
excited band in $^{151}$Tb, denoted by $^{152}$Dy(1) and
$^{151}$Tb(2), respectvely.  Details of the calculation methods
can be found in Ref.\spc\cite{J2D2}.  The coupling constants of
the energy-density functional have been determined from the
parameters of three Skyrme forces, SkM* \cite{BQB82}, SkP
\cite{DFT84}, and SIII \cite{Bei75}. The left-hand-side panels
show the results obtained for the complete functionals, while
those at the right-hand-side correspond to omitted time-odd
terms ${\cal H}^{\text{odd}}(\bbox{r})$$\equiv$0, i.e., to
   \begin{equation}\label{eq313}
   C^{s}=C^{\Delta{s}}=C^{T}=C^{j}=C^{\nabla{j}}=0.
   \end{equation}
For the $^{151}$Tb(1) and $^{150}$Gd(2) bands, the analogous
results are shown in Fig.\spc\ref{fig02}.

The omission of time-odd terms in the Hartree-Fock approach is analogous
to the standard phenomenological rotating mean-field approach
where the selfconsistent changes of the mean field due to
rotations are not taken into account.  Results of latter type
calculations, for the same SD bands as considered in the present
study, can be found in Refs.\spc\cite{NWJ89} and \cite{Rag93}.

For the complete Skyrme functionals, the dynamic moments $\jde$
of the $^{152}$Dy(1) band obtained for SkM*, SkP, and SIII
interactions are very similar, see part (a) in the
left-hand-side panel of Figs.\spc\ref{fig01} and \ref{fig02}.
This is not the case when the time-odd terms are omitted, as
show in the corresponding right-hand-side panels.  For SkM* and
SIII, the omission of the time-odd terms significantly decreases
the values of $\jde$, while for SkP the changes of $\jde$ are
much smaller. This feature is related to the values of the
isoscalar nuclear-matter effective mass, which for SkP is equal
to the free nucleon mass, $m^*$=$m$, while for SkM* and SIII is
smaller by the factor of 0.79 and 0.76, respectively.  As a
consequnce, for SkP the isoscalar coupling constant $C^j_0$ is
equal to zero \cite{J2D2}, and the modifications introduced by
omitting the time-odd terms, Eq.\spc(\ref{eq313}), do not
influence $\jde$ in a very strong way. On the other hand,
whenever this coupling constant is relatively large, as is for
SkM* and SIII, the corresponding changes in $\jde$ are well
prononced.

These results illustrate the importance for the rotational
properties of the nucleus of the $\bbox{j}^2$ term in the energy
density. The cranking term mainly induces the nonzero flow of
nuclear matter, as given by the current density $\bbox{j}$, whch
influences the time-odd nuclear mean field provided the
corresponding coupling constant $C^j$ is nonzero.

In spite of the strong influence on the values of $\jde$, the
omission of the time-odd terms has a very small effect on the
sameness of $\jde$ in pairs of identical bands. In parts (b) of
Figs.\spc\ref{fig01} and \ref{fig02} we show the the relative
dynamical moments $\delta\jde$, i.e., the differences between
the values of $\jde$ calculated for two pairs of identical bands
$^{151}$Tb(2)--$^{152}$Dy(1) and $^{150}$Gd(2)--$^{151}$Tb(1).
One can see that the omission of the time-odd terms influences the
values of $\jde$ in a very similar way for both members of each
pair. Even if the values may change by as much as 15\,{\umi},
the relative values are always well below 2\,{\umi}. This shows
that the sameness of $\jde$ is not governed by the time-odd
terms in the mean-fields, but rather can be attributed to
general geometric properties of the uderlying orbital {\ptzzu}.

A different patterns is obtained for relative alignments
shown in parts (c) of the figures. Here the omision of the
time-odd terms leads to important changes of the
relative alignments, and also different relative alignments
are obtained of the three Skyrme forces studied.

In Figs.\spc\ref{fig03} and \ref{fig04} we present a detailed
analysis of the time-odd terms appearing in the energy density
for the SkP interaction.  As shown in the legends, the results
presented in the left-hand-side panels correspond to the
following conditions:  (i) the complete functional, (ii) $C_t^{
T}$=$C_t^{   J}$=0, (iii) $C_t^{\Delta s}$=$C_t^{T}$=$C_t^{
J}$=0, and then (iv) $C_t^{ s}$=$C_t^{\Delta s}$=$C_t^{
T}$=$C_t^{   J}$=0.  In the right-hand-panels, in addtion to the
last of these conditions we set to zero either one, or both of
the $C^j$ and $C^{\nabla{j}}$ coupling constants.

As discussed above, the value of the effective mass $m^*$=$m$
renders the dynamical moments $\jde$, the relative dynamical
moments, and the relative alignments $\delta{I}$ almost
independent of the $C^j$ coupling constants. (A weak residual
dependence results from a non-zero value of the isovector
coupling constant $C^j_1$, see Ref.\spc\cite{J2D2}.) On the
other hand, the values of the relative alingments strongly
depend on the $C^{\nabla{j}}$ coupling constants, and are
closest to the experimental data when these coupling constants
are equal to zero.

\section{Conclusions}
\label{sec4}

In the present study we have presented a systematic construction
of the energy-density functional in the frame of the nuclear
local density approximation. Such a construction is based on a
few simple prescriptions and leads to the functional identical to that
obtained by using the Skyrme effective interaction. The only
exception is the term given by the square of the divergence of
the spin density, which can only be obtained if the tensor
component is added to the Skyrme interaction.

Following Ref.\spc\cite{J2D2}, we have here presented the
analysis of the time-odd components of the mean field of
rotating superdeformed nuclei. Special attention has been
devoted to properties of the SkP Skyrme interaction which is
characterized be the effective mass $m^*$=$m$. In particular, we
have shown that the sameness of the dynamical moments $\jde$
{\em does not depend} on whether the time-odd terms are, or are
not taken into account.  On the other hand, the sameness of the
corresponding alignments {\em does depend} on these time-odd
terms.  Therefore, one may expect that a systematic
investigation of the rotational alingments may serve as a
tool for establishing the properties of time-odd component of
the rotating nuclear mean field.

\renewcommand{\topfraction}{0.0}
\renewcommand{\bottomfraction}{0.0}
\renewcommand{\floatpagefraction}{0.0}
\setcounter{topnumber}{20}
\setcounter{bottomnumber}{20}
\setcounter{totalnumber}{20}

\begin{figure}[ht]
\caption[F]{%
Calculated dynamical moment ${{\cal J}^{(2)}}$ for the yrast
band of $^{152}$Dy, part (a), the relative dynamical moment
$\delta{{\cal J}^{(2)}}$ calculated for the $^{151}$Tb(2) and
$^{152}$Dy(1) bands, part (b), and the relative alignment
$\delta I$ between these two bands, part (c). Left-hand-side
panels show the results for complete Skyrme functionals of the SkM*,
SkP, and SIII interactions, and the right-hand-side panels show
the results with omitted time-odd terms. The experimental points
are denoted by asterisks.  Note the scale in (b) expanded five
times as compared to (a).}
\label{fig01}
\end{figure}

\begin{figure}[ht]
\caption[F]{%
Same as Fig.~\protect\ref{fig01}, but for the $^{151}$Tb(1) band, (a),
and for the differences between the $^{150}$Gd(2) and
$^{151}$Tb(1) bands, (b)
and (c).}
\label{fig02}
\end{figure}

\begin{figure}[ht]
\caption[F]{%
Similar as in Fig.~\protect\ref{fig01}, but for different
time-odd terms omitted in the energy density, see text.}
\label{fig03}
\end{figure}

\begin{figure}[htb]
\caption[F]{%
Same as Fig.~\protect\ref{fig03}, but for the $^{151}$Tb(1) band, (a),
and for the differences between the $^{150}$Gd(2) and
$^{151}$Tb(1) bands, (b)
and (c).}
\label{fig04}
\end{figure}

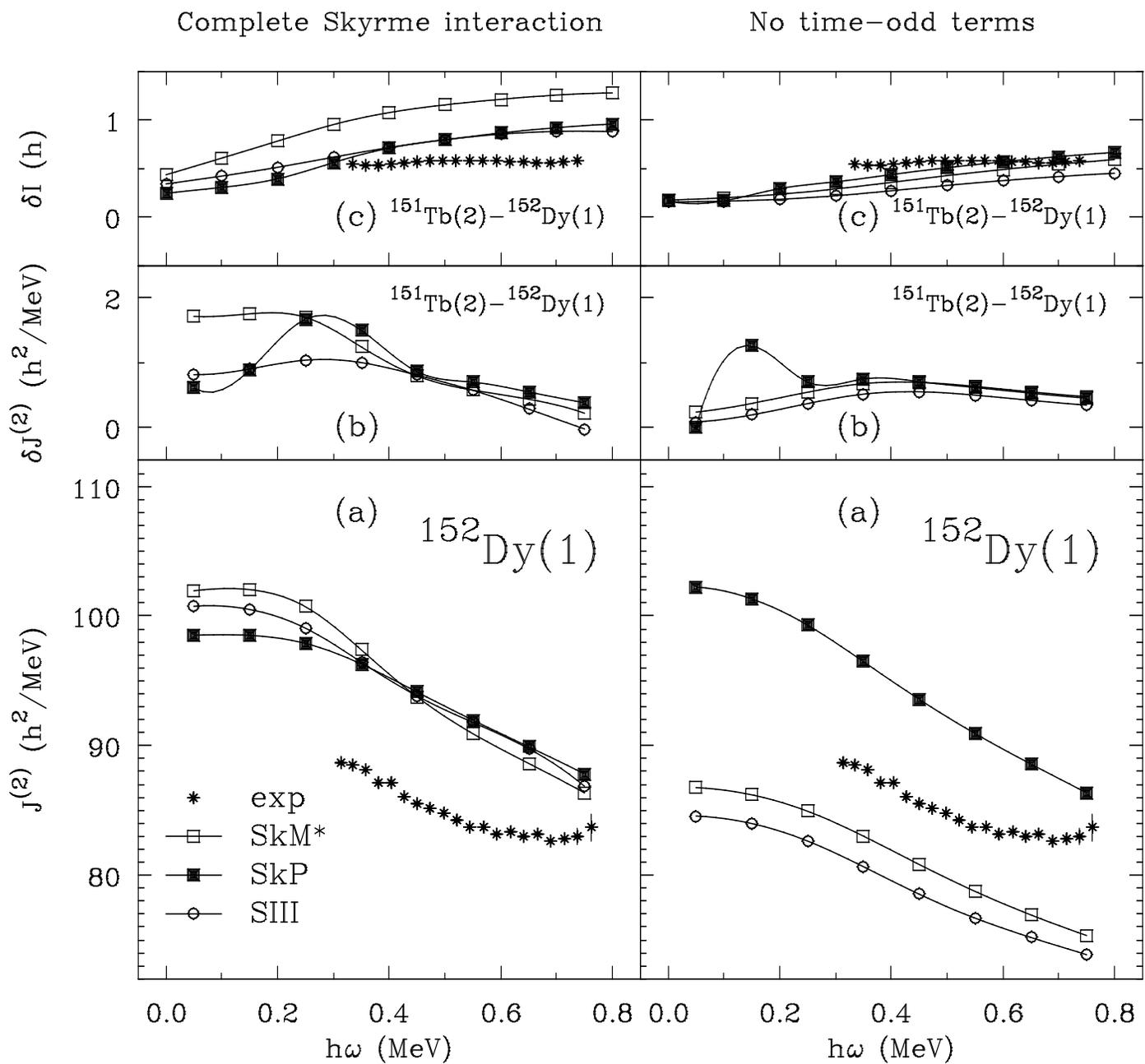
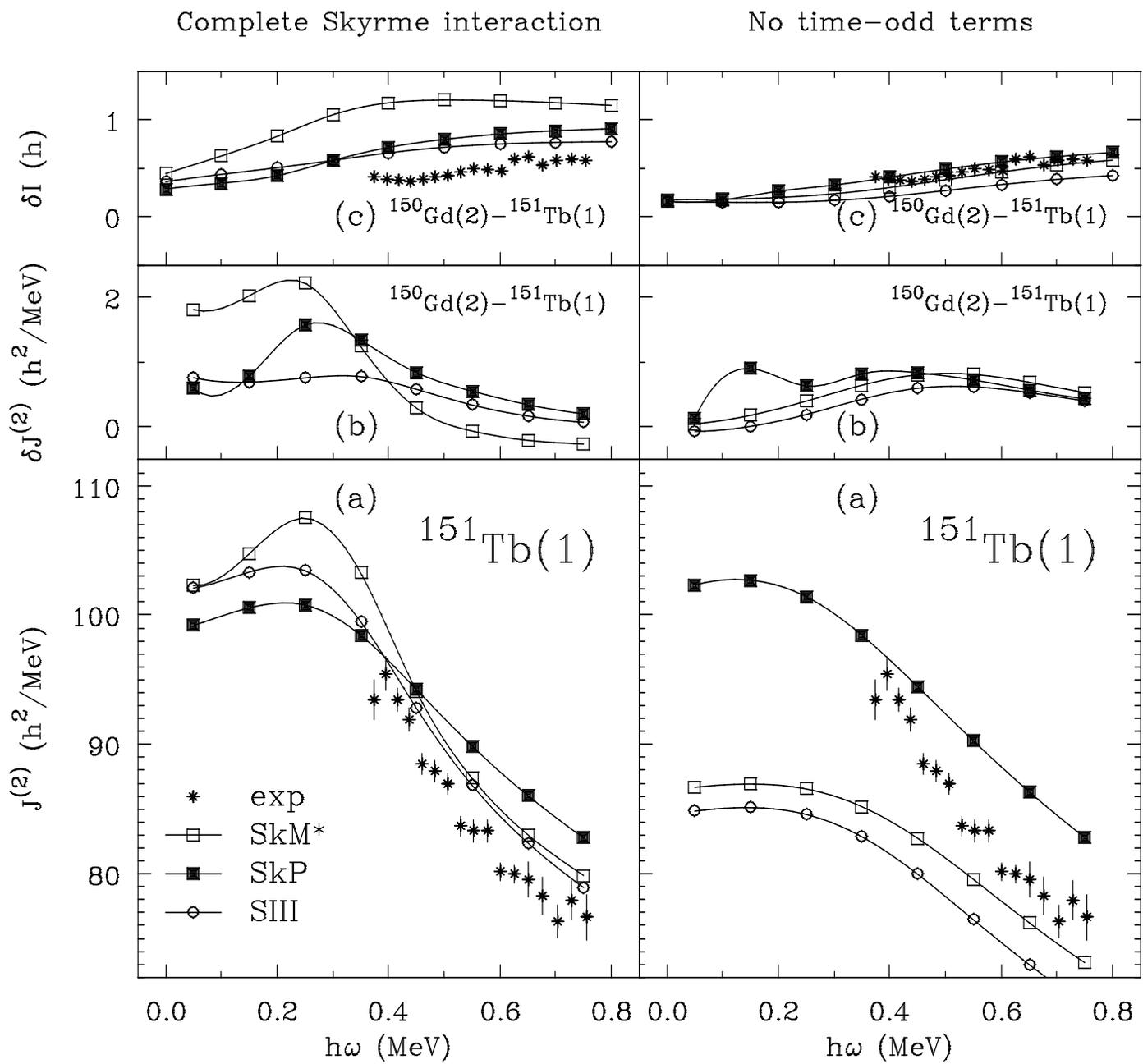
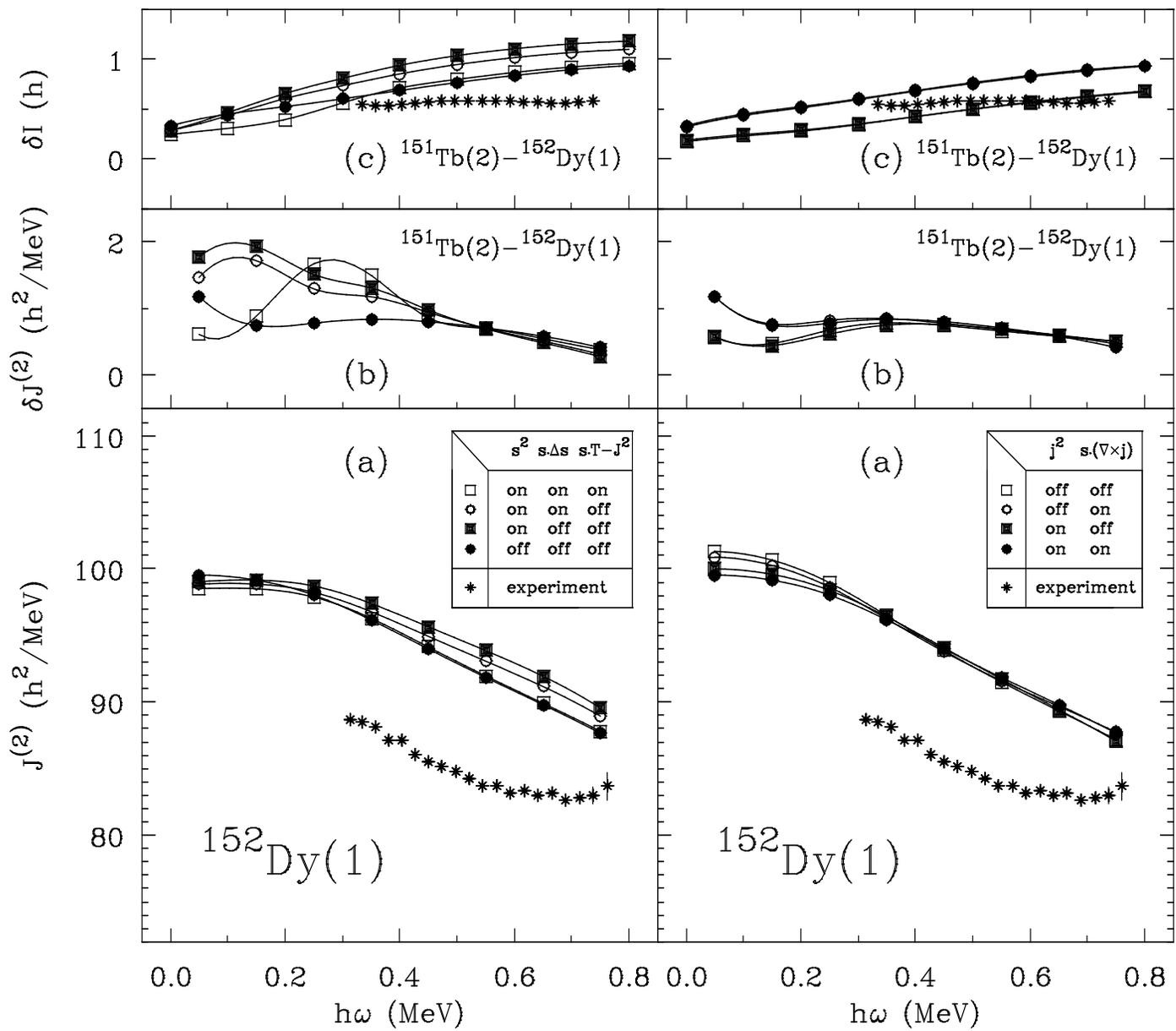
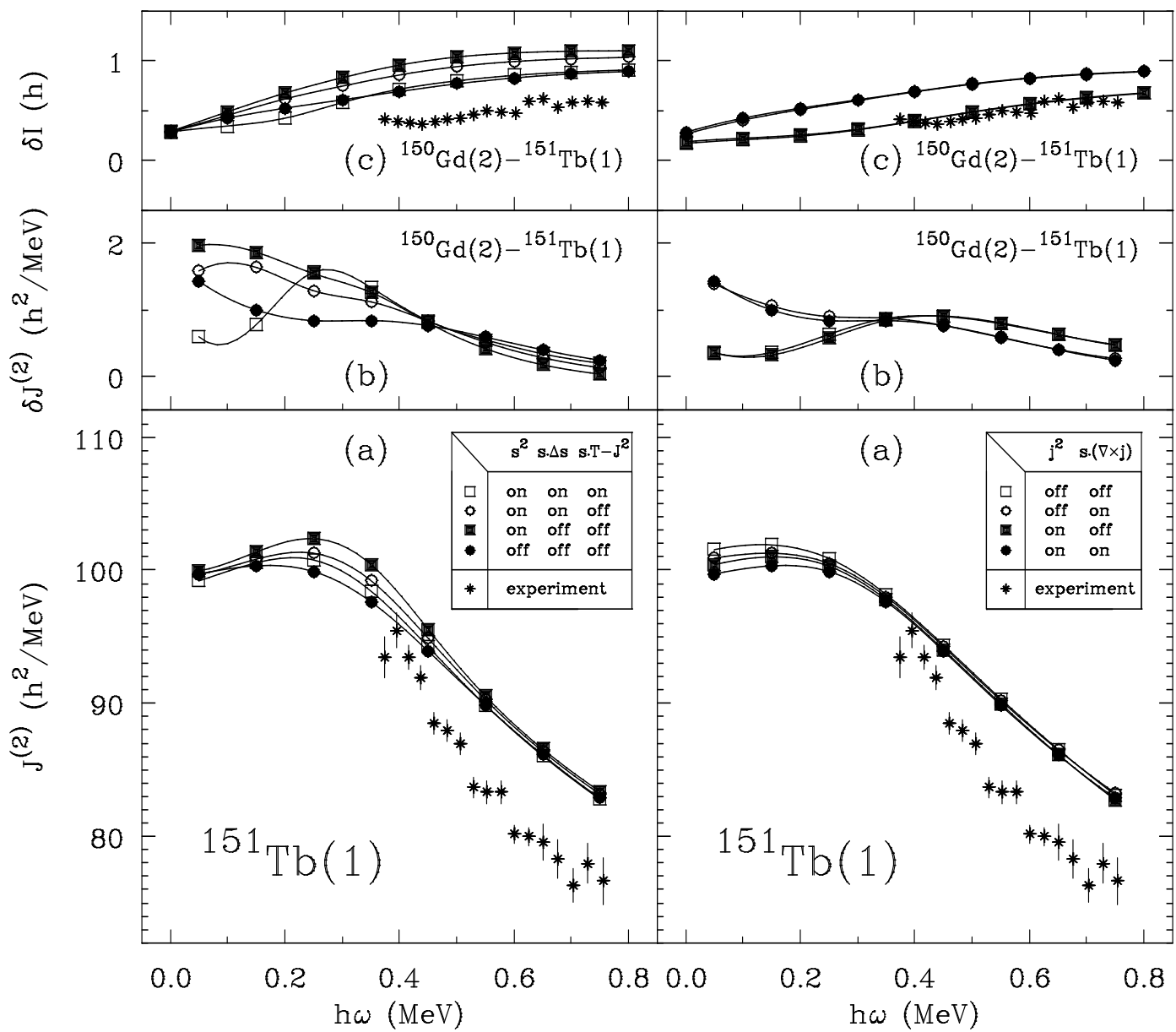
\end{document}